%% file: CrII.tex
\documentclass[twocolumn,aps,pra,showpacs]{revtex4-1}

\usepackage{graphicx}
\usepackage{amsmath}
\usepackage{relsize}
\usepackage{dcolumn}

\include{phys_jdf}

\newcommand{\CrII}{Cr\,II}
\newcommand{\ZnII}{Zn\,II}
\newcommand{\SrII}{Sr\,II}
\newcommand{\FeII}{Fe\,II}
\newcommand{\MgII}{Mg\,II}

\newcommand{\knms}{\ensuremath{k_\textrm{\relsize{-1}{NMS}}}}
\newcommand{\ksms}{\ensuremath{k_\textrm{\relsize{-1}{SMS}}}}

\newcolumntype{d}{D{.}{.}{-1}}   
\newcolumntype{b}{D{(}{\,(}{-1}} 

\newcommand{\cm}{\ensuremath{\textrm{cm}^{-1}}}
\newcommand{\GHzamu}{GHz$\cdot$amu}
\newcommand{\ms}{\ensuremath{\textrm{m\,s}^{-1}}}

\newcommand{\bra}[1]{\ensuremath{\left< #1 \right|}}
\newcommand{\ket}[1]{\ensuremath{\left| #1 \right>}}
\newcommand{\rtw}{\ensuremath{\rightarrow}}
\newcommand{\vect}[1]{\mathbf{#1}}

\newcommand{\SigOne}{\ensuremath{\Sigma^{(1)}}}
\newcommand{\SigTwo}{\ensuremath{\Sigma^{(2)}}}
\newcommand{\SigThree}{\ensuremath{\Sigma^{(3)}}}

\newcommand{\VN}{\ensuremath{V^{N}}}
\newcommand{\VNm}{\ensuremath{V^{N-1}}}
\newcommand{\GS}{\ensuremath{3d^5\ ^6\!S_{5/2}}}

\newcommand{\Cite}[1]{\mbox{Ref.~\cite{#1}}}

\newcommand{\eref}[1]{(\ref{#1})}

\newcommand{\Tref}[1]{Table~\ref{#1}}
\newcommand{\Fig}[1]{Fig.~\ref{#1}}
\newcommand{\Sec}[1]{Section~\ref{#1}}

\begin{document}

\title{Isotope shifts and relativistic shifts of \CrII\ for study of $\boldsymbol{\alpha}$-variation in quasar absorption spectra}

\author{J. C. Berengut}
\affiliation{School of Physics, University of New South Wales, Sydney, NSW 2052, Australia}

\date{11 October 2011}

\pacs{31.30.Gs, 31.15.am, 95.30.Ky, 06.20.Jr}

\begin{abstract}
We use the combination of configuration interaction and many-body perturbation theory method (CI+MBPT) to perform \emph{ab initio} calculations the low-energy spectra of \CrII\ with high accuracy. It is found that second-order MBPT diagrams should be included in a consistent and complete way for the MBPT to improve the accuracy of calculations in this five-valence-electron system. This contrasts with previous ions with fewer valence electrons where it was found that single-valence-electron diagrams dominate the corrections. Isotope shifts and relativistic shifts ($q$-values) are calculated for use in astronomical determination of the fine-structure constant in quasar absorption spectra.
\end{abstract}

\maketitle

\section{Introduction}

Quasar absorption systems provide a unique probe of the value of fundamental constants throughout much of the visible Universe. The many-multiplet (MM) method enables the most complete analysis of optical spectra in the search for space-time variation of the fine-structure constant, $\alpha = e^2/\hbar c$~\cite{dzuba99prl,dzuba99pra}. It makes use of all transitions seen in all ions in a given quasar absorption system to gain statistical significance and control systematics. Early results using spectra taken from the Keck telescope suggested that $\alpha$ may have been smaller in the past~\cite{webb99prl,murphy03mnras,murphy04lnp}, however when combined with new systems observed with the Very Large Telescope (VLT) the data is more consistent with a spatial variation in the fine-structure constant~\cite{webb10arxiv}. The gradient in values of $\alpha$ reconciles all existing measurements of $\alpha$-variation~\cite{berengut11jpcs}. In particular the early Keck results that indicated a constant offset or ``monopole'' model, are entirely consistent with the spatial gradient ``dipole'' model since Keck mainly sees in the northern hemisphere (the $\alpha$-dipole axis is oriented $\sim 30^\circ$ from the equatorial axis). By contrast the VLT data is taken mainly in the southern sky.

A spatial variation of $\alpha$ would manifest itself in a variety of terrestrial~\cite{berengut10arxiv0} and astrophysical~\cite{berengut11prd} systems, which could be used to confirm the dipole. It is also possible to devise complementary tests using subsets of the quasar absorption system data which may involve different systematics. One such test, currently underway, is a variant of the many-multiplet method that only uses transitions in \CrII\ and \ZnII~\cite{malec12unpublished}. The transitions have opposite $\alpha$-sensitivities and so a comparison of them is very sensitive to $\alpha$-variation: \ZnII\ transitions are $s-p$ and hence their frequency increases if $\alpha$ increases, while the \CrII\ transitions are $d-p$ so their frequency decreases with increasing values of $\alpha$. Furthermore the transitions are very close in energy. This means that only a small part of the optical spectrum is analysed, resulting in different (perhaps smaller) systematics. Of particular concern are ``intra-order shifts'': velocity shifts of unknown origin within each echelle order in the spectrograph~\cite{griest10apj,whitmore10apj}. This systematic may differently affect measurements of $\alpha$-variation when only \CrII\ and \ZnII\ lines are utilized, compared to studies where a larger number and wider variety of transitions are used.

One problem with using \CrII\ and \ZnII\ transitions exclusively is that they are weak. Of course, this is the reason why they don't play a major role in the full MM analysis despite being included whenever available. However there exist certain quasar absorption systems in which \ZnII\ lines are particularly strong~\cite{herbert-fort06ajpac}, and from these ``metal strong'' systems can be drawn a relatively large sample with which to perform the \CrII/\ZnII\ analysis.

One potential systematic that has plagued all MM analyses is isotope abundance~\cite{murphy01mnrasB,berengut03pra,kozlov04pra}. Isotopic structure cannot be resolved in the absorption spectra, so generally terrestrial isotopic abundances are assumed for the absorber. Any deviation from terrestrial abundances would shift the centroid of the line profile, and this might mimic a change in $\alpha$. Even quantifying the systematic can be difficult because the isotopic structures themselves are unknown for many of the UV transitions used in the MM analysis. The systematic is lessened in the context of a dipole result, since the isotope abundances would need to vary according to direction in the sky, which in itself would violate cosmological isotropy. Nevertheless, in order to quantify possible systematics the isotope structure should be known for all transitions used in the analysis, hence considerable efforts by many groups to calculate and measure them~(see, e.g.~\cite{safronova01pra,korol07pra,salumbides06mnras,porsev09pra0,berengut10hypint}).

In this paper we calculate the isotope shifts and relativistic shifts of the \CrII\ transitions seen in quasar absorption spectra. The corresponding parameters for \ZnII\ have been calculated previously~\cite{berengut03pra,dzuba07pra}. Our final results are presented in Tables~\ref{tab:final_q} and~\ref{tab:final_IS}.

\section{Method}
\label{sec:method}

The \emph{ab initio} CI+MBPT method~\cite{dzuba96pra} is described in full elsewhere~\cite{berengut06pra}. Details of relevance for our \CrII\ calculation are presented below.

\subsection{Energy calculation}
\label{sec:energy}

Any perturbative theory works best when the perturbations are as small as possible. In \CrII\ the $d$-wave electrons play an important role in shaping the atomic core, and so they should be included in the initial approximation. As in previous works, our single-particle wavefunctions are calculated using Dirac-Fock (relativistic Hartree-Fock). We explore two Dirac-Fock configurations: \VN, which includes a half-filled $3d^5$ subshell; and \VNm, which includes $3d^4$. In both cases we simply scale the potential due to the filled $3d$ subshell by the number of electrons to provide a ``configuration averaged'' initial wavefunction. The choice of starting approximation is essential to obtaining a good final spectrum for \CrII, but also leads to potentially large subtraction diagrams in many-body perturbation theory, as will be demonstrated.

Once we have a Dirac-Fock potential for the core, we diagonalize the Dirac-Fock Hamiltonian over a set of 40 $B$-splines~\cite{johnson86prl} spanning 40 atomic units to obtain a large set of valence and virtual orbitals from which we select those with the lowest eigenvalues. A set of configurations of valence electrons \ket{I} are generated, from which eigenfunctions of the complete Coulomb-potential Hamiltonian are calculated. We find that almost complete convergence of the CI calculation can be obtained using the basis $20spdf$: that is we use $s$-wave states labeled 1 -- 20, $p$-wave states labeled 2 -- 20, etc. (For the lowest eigenvalue states the label is just the principal quantum number.) With 5 valence electrons it is not possible to include all configurations and we must select those that contribute most to the wavefunction. We include all configurations that can be formed by one-particle excitations from the leading configurations $3d^5$, $3d^44s$, and $3d^44p$, as well as two-particle excitations from these same configurations up to the $6sp9d6\!f$ orbitals. The effects of higher orbitals and three-particle excitations were found to be small and were not included.

Having achieved high saturation of the CI calculation, core-valence effects are included using second-order MBPT by modifying matrix elements of the Hamiltonian:
\begin{equation}
\label{eq:CI+MBPT}
  H_{IJ} \rtw H_{IJ} + \sum_M \frac{\bra{I}H\ket{M}\bra{M}H\ket{J}}{E-E_M}\,,
\end{equation}
where the states \ket{M} include all Slater determinants that have core excitations. The MBPT sum may be further separated into one-, two-, and three-valence-electron parts, denoted \SigOne, \SigTwo\ and \SigThree\ in Refs.~\cite{berengut06pra,berengut08jpb}. Goldstone diagrams and analytical expressions for these are presented in~\cite{berengut06pra}. The states \ket{M} include excitations from all core states into virtual states up to $30spdf\!g$. The effects on the energy calculation of including \SigOne, \SigTwo, and \SigThree\ sequentially are shown in Tables~\ref{tab:energy} and~\ref{tab:energyVN} (for the \VNm\ and \VN\ starting approximations, respectively).

\begin{table}[tb]
\caption{\label{tab:energy} Energy spectrum of \CrII\ with orbitals calculated in the \VNm\ approximation, relative to the experimentally determined ground state, \GS. Successive additions of \SigOne, \SigTwo, and \SigThree\ are shown. (Energies in \cm.)}
\begin{ruledtabular}
\begin{tabular}{lrrrrrr}
Level & J & \multicolumn{1}{c}{CI} & \multicolumn{1}{c}{+\SigOne} & \multicolumn{1}{c}{+\SigTwo} & \multicolumn{1}{c}{+\SigThree} & \multicolumn{1}{c}{Expt.~\cite{NIST}} \\
\hline
$3d^4 4s\ ^6\!D$   & 5/2 & 8505 & $-7294$ & 6888 & 9491 & 12148 \\
                   & 7/2 & 8682 & $-7128$ & 7095 & 9699 & 12304 \\
$3d^4 4p\ ^6\!F^o$ & 1/2 & 41720 & 27542 & 40933 & 44533 & 46823 \\
                   & 3/2 & 41805 & 27635 & 41028 & 44638 & 46905 \\
                   & 5/2 & 41945 & 27788 & 41184 & 44798 & 47040 \\
                   & 7/2 & 42140 & 28000 & 41401 & 45036 & 47227 \\
                   & 9/2 & 42387 & 28271 & 41676 & 45320 & 47465 \\
                   &11/2 & 42687 & 28600 & 42008 & 45668 & 47752 \\
$3d^4 4p\ ^6\!P^o$ & 3/2 & 43214 & 29890 & 42534 & 46074 & 48399 \\
                   & 5/2 & 43333 & 29982 & 42686 & 46216 & 48491 \\
                   & 7/2 & 43515 & 30132 & 42915 & 46420 & 48632 \\
$3d^4 4p\ ^4\!P^o$ & 1/2 & 43793 & 30108 & 42725 & 46584 & 48749 \\
                   & 3/2 & 44057 & 30328 & 43038 & 46907 & 49006 \\
                   & 5/2 & 44787 & 30674 & 43880 & 47717 & 49706 \\
$3d^4 4p\ ^6\!D^o$ & 1/2 & 44594 & 29189 & 43659 & 47460 & 49493 \\
                   & 3/2 & 44664 & 29323 & 43741 & 47552 & 49565 \\
                   & 5/2 & 44451 & 29510 & 43492 & 47330 & 49352 \\
                   & 7/2 & 44840 & 29721 & 43959 & 47727 & 49646 \\
                   & 9/2 & 45054 & 29940 & 44209 & 47994 & 49838 \\
\end{tabular}
\end{ruledtabular}
\end{table}

\begin{table}[tb]
\caption{\label{tab:energyVN} Energy spectrum of \CrII\ with orbitals calculated in the \VN\ approximation, relative to the experimentally determined ground state, \GS. Successive additions of \SigOne, \SigTwo, and \SigThree\ are shown. (Energies in \cm.)}
\begin{ruledtabular}
\begin{tabular}{lrrrrrr}
Level & J & \multicolumn{1}{c}{CI} & \multicolumn{1}{c}{+\SigOne} & \multicolumn{1}{c}{+\SigTwo} & \multicolumn{1}{c}{+\SigThree} & \multicolumn{1}{c}{Expt.~\cite{NIST}} \\
\hline
$3d^4 4s\ ^6\!D$   & 5/2 & 6688 & $-8662$ & 3953 & 6415 & 12148 \\
                   & 7/2 & 6862 & $-8515$ & 4157 & 6617 & 12304 \\
$3d^4 4p\ ^6\!F^o$ & 1/2 & 41540 & 26023 & 38794 & 41627 & 46823 \\
                   & 3/2 & 41626 & 26110 & 38888 & 41726 & 46905 \\
                   & 5/2 & 41768 & 26257 & 39044 & 41887 & 47040 \\
                   & 7/2 & 41966 & 26461 & 39260 & 42114 & 47227 \\
                   & 9/2 & 42217 & 26724 & 39534 & 42399 & 47465 \\
                   &11/2 & 42521 & 27047 & 39865 & 42747 & 47752 \\
$3d^4 4p\ ^6\!P^o$ & 3/2 & 43124 & 27918 & 40888 & 43534 & 48399 \\
                   & 5/2 & 43242 & 27987 & 41041 & 43674 & 48491 \\
                   & 7/2 & 43421 & 28113 & 41265 & 43885 & 48632 \\
$3d^4 4p\ ^4\!P^o$ & 1/2 & 43816 & 28030 & 41092 & 43995 & 48749 \\
                   & 3/2 & 44033 & 28307 & 41320 & 44246 & 49006 \\
                   & 5/2 & 44812 & 29073 & 42251 & 45175 & 49706 \\
$3d^4 4p\ ^6\!D^o$ & 1/2 & 44432 & 28788 & 41749 & 44708 & 49493 \\
                   & 3/2 & 44552 & 28878 & 41915 & 44861 & 49565 \\
                   & 5/2 & 44303 & 28603 & 41591 & 44537 & 49352 \\
                   & 7/2 & 44543 & 28849 & 41851 & 44815 & 49646 \\
                   & 9/2 & 44766 & 29041 & 42108 & 45078 & 49838 \\
\end{tabular}
\end{ruledtabular}
\end{table}

\subsection{Isotope shift and relativistic shift}
\label{sec:isotope}

Isotope shifts in atomic transition frequencies come from two sources: the finite size of the nucleus (field shift), and the recoil of the nucleus (mass shift). This mass shift is usually divided into the normal mass shift (NMS), which is easily calculated from the transition frequency, and the specific mass shift (SMS). The mass shift is more important for light elements, while for heavy elements the field shift dominates. In the case of \CrII, the field shift is small; this paper is concerned with the mass-shift contribution, which is more difficult to calculate. The difference in the transition frequency, $\omega$, between an isotope with mass number~$A'$ and an isotope~$A$, $\delta \omega^{A',A} = \omega^{A'} - \omega^A$, can be expressed as~\cite{berengut03pra}
\begin{equation}
\label{eq:IS}
\delta \omega^{A',A} = (\knms + \ksms) \left(\frac{1}{A'} - \frac{1}{A}\right)
  + F \delta\!\left< r^2 \right>^{A', A} \,,
\end{equation}
where $\delta\!\left< r^2 \right>$ is the change in mean-square nuclear charge radius.
The normal mass shift constant can be expressed  (in atomic units $\hbar = e = m_e = 1$)
\[
  \knms = \frac{1}{2m_u} \sum_i p_i^2 = -\frac{\omega}{m_u} \,,
\]
where $m_u = 1823$ is the ratio of the atomic mass unit to the electron mass, and the sum is over all electron momenta, $\vect{p}_i$. The specific-mass-shift constant
\[
  \ksms = \frac{1}{m_u} \sum_{i<j} \vect{p}_i \cdot \vect{p}_j
\]
and field-shift constant $F$ are more difficult to calculate. We use the non-relativistic form of the mass-shift operator; relativistic corrections for optical transitions in light atoms are on the order of few percent and can be neglected~\cite{korol07pra}.

To calculate \ksms\ we use the all-order finite-field scaling method. Here a rescaled two-body SMS operator is added to the Coulomb potential everywhere that it appears in an energy calculation:
\begin{equation}
\label{eq:tilde_Q}
\tilde{Q} = \frac{1}{\left| \vect{r}_1 - \vect{r}_2\right|} + \lambda \vect{p}_1 \cdot \vect{p}_2 \,.
\end{equation}
We recover the specific-mass-shift constant as
\begin{equation}
\ksms = \frac{d\omega}{d\lambda}\bigg{|}_{\lambda = 0}\,.
\end{equation}
The operator $\tilde{Q}$ has the same symmetry and structure as the Coulomb operator (see Appendix~A of Ref.~\cite{berengut06pra}). We have previously shown that good agreement with experimental isotope shift can be obtained in many-valence-electron atoms and ions by using this finite-field in a CI+MBPT energy calculation~\cite{berengut04praA,berengut05pra,berengut06pra,berengut08jpb}.

The relativistic shift of a transition may be calculated in a similar fashion. We simply recalculate the transition energies, $\omega$, from the very beginning using different values of $\alpha$ from the laboratory value $\alpha_0$. The sensitivity to variation of the fine-structure constant is then extracted using
\begin{equation}
q = \frac{d\omega}{d\alpha^2}\bigg{|}_{\alpha = \alpha_0} \,.
\end{equation}

\section{Results and discussion}
\label{sec:results}

\CrII\ has five valence electrons, and these have a significant impact on the form of the basis orbitals. For the CI+MBPT method to work well it is important to have good initial orbitals, and so our Dirac-Fock and subsequent $B$-spline codes include the $3d^4$ or $3d^5$ configuration in the core, as described in \Sec{sec:energy}. (Our calculations show that saturation of the CI can be met satisfactorily in both \VNm\ and \VN\ approximations.) In the CI+MBPT code, the $3d$ orbitals are then stripped from the core and become valence orbitals for the purposes of both the CI and MBPT components of the calculation. In this way excitations from the $3d$ shell are treated non-perturbatively.

\begin{table}[tb]
\caption{\label{tab:sms} Isotope shift constants \knms\ and \ksms\ for transitions to the ground state \GS\ (\GHzamu). }
\begin{ruledtabular}
\begin{tabular}{lrrrrrr}
Level & \multicolumn{1}{c}{$J$} & \multicolumn{1}{c}{\knms} & \multicolumn{2}{c}{\ksms\ (CI)} & \multicolumn{2}{c}{\ksms\ (CI+$\Sigma$)} \\
      &   & & \multicolumn{1}{c}{\VN} & \multicolumn{1}{c}{\VNm} & \multicolumn{1}{c}{\quad\VN} & \multicolumn{1}{c}{\VNm} \\ 
\hline
$3d^4 4s\ ^6\!D$   & 5/2 & $-200$ & $4520$ & $4326$ & $4944$ & $5062$ \\
                   & 7/2 & $-202$ & $4532$ & $4337$ & $4950$ & $5069$ \\
$3d^4 4p\ ^6\!F^o$ & 1/2 & $-770$ & $3964$ & $3398$ & $4303$ & $4127$ \\
                   & 3/2 & $-771$ & $3970$ & $3403$ & $4314$ & $4140$ \\
                   & 5/2 & $-774$ & $3980$ & $3412$ & $4330$ & $4161$ \\
                   & 7/2 & $-777$ & $3992$ & $3423$ & $4346$ & $4178$ \\
                   & 9/2 & $-781$ & $4009$ & $3438$ & $4359$ & $4191$ \\
                   &11/2 & $-785$ & $4028$ & $3456$ & $4362$ & $4188$ \\
$3d^4 4p\ ^6\!P^o$ & 3/2 & $-796$ & $4072$ & $3555$ & $4189$ & $4161$ \\
                   & 5/2 & $-797$ & $4080$ & $3563$ & $4143$ & $4090$ \\
                   & 7/2 & $-800$ & $4093$ & $3575$ & $4203$ & $4180$ \\
$3d^4 4p\ ^4\!P^o$ & 1/2 & $-802$ & $3981$ & $3495$ & $4227$ & $4068$ \\
                   & 3/2 & $-806$ & $3983$ & $3507$ & $4257$ & $4187$ \\
                   & 5/2 & $-817$ & $4071$ & $3493$ & $4271$ & $4188$ \\
$3d^4 4p\ ^6\!D^o$ & 1/2 & $-814$ & $4016$ & $3458$ & $4195$ & $4137$ \\
                   & 3/2 & $-815$ & $4037$ & $3467$ & $4267$ & $4156$ \\
                   & 5/2 & $-812$ & $3982$ & $3511$ & $4278$ & $4183$ \\
                   & 7/2 & $-816$ & $3989$ & $3451$ & $4280$ & $4129$ \\
                   & 9/2 & $-820$ & $4003$ & $3463$ & $4302$ & $4153$ \\
\end{tabular}
\end{ruledtabular}
\end{table}

To calculate the MBPT diagrams one must include the change in effective core-potential, \mbox{$V^{N-5}-\VN$} or \mbox{$V^{N-5}-\VNm$} depending on the initial approximation. MBPT diagrams that include this interaction are known as subtraction diagrams, and in our calculation they are huge. There are three subtraction diagrams in \SigOne\ and two in \SigTwo\ (these are shown in Figs.~2 and~4 of \cite{berengut06pra}, respectively). When the subtraction diagrams in \SigOne\ are included, they significantly and adversely affect the energies obtained, as can be seen in the CI+\SigOne\ columns (labelled ``+\SigOne'') of Tables~\ref{tab:energy} and~\ref{tab:energyVN}. However, it turns out that these adverse effects are nearly completely compensated when \SigTwo\ and \SigThree\ are also included. When all second-order diagrams are included consistently the energies and wavefunctions are improved by the addition.

One might consider what happens if the subtraction diagrams are simply neglected from the calculation. Indeed the CI+\SigOne\ energies are improved. However when \SigTwo\ is added (either with or without the two-valence-electron subtraction diagrams) the energy levels obtained are again in very poor agreement with experiment. Thus it is not only inconsistent to leave out the subtraction diagrams, but it gives very poor results when all other second-order MBPT terms are included.

\begin{figure}[tb]
\caption{\label{fig:diagrams} Two large diagrams that affect the ground state $3d^5$ multiplet that partially cancel. The labels refer to their designation in~\cite{berengut03pra}.}
\includegraphics[width=8cm]{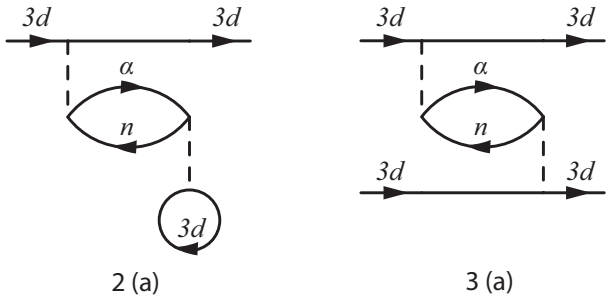}
\end{figure}

The behaviour can be explained by examining the form of the \SigOne\ and \SigTwo\ diagrams. For example, consider Fig.~2(a) and Fig.~3(a) from~\cite{berengut06pra} with all external lines representing $3d$ electrons (see~\Fig{fig:diagrams}). The subtraction diagram 2(a) has opposite sign to the zero multipole ($k=0$) part of 3(a). This kind of cancellation is what finally suppresses the large subtraction diagrams, and is the reason why all second-order $\Sigma$ diagrams must be included consistently.

The energies obtained are slightly better in the \VNm\ calculation. However, consider our calculations of the isotope shift constant \ksms, presented in \Tref{tab:sms}. We find that the pure CI calculations give different results in the \VN\ and \VNm\ approximations, yet when the core-valence interactions are included self-consistently the agreement is much improved: the disagreement is reduced from $\sim 13\%$ to less than $4\%$ and in the $3d^4 4p\ ^6\!P^o$ transitions of astrophysical interest, more like $\sim 1\%$.

The calculation of $q$-values follow the same trend (\Tref{tab:qvalue}). In this case the results are far less sensitive to details of the wavefunction: instead the relativistic effects are determined by the form of the wavefunctions near the nucleus. We see very strong agreement between our \VN\ and \VNm\ results, especially after all $\Sigma$ diagrams are included consistently. In the transitions of astrophysical interest, the different starting approximations leads to disagreements of the order $\sim 25\%$ in the pure CI case but $\sim 5\%$ when MBPT is included.

\begin{table}[tb]
\caption{\label{tab:qvalue} $g$-factors and relativistic shifts, $q$ (\cm), for transitions to the ground state \GS. Experimental $g$-factors are taken from \Cite{NIST}; calculated values are for the full CI+$\Sigma$ method in the \VNm\ approximation.}
\begin{ruledtabular}
\begin{tabular}{lrrrrrrr}
Level & \multicolumn{1}{c}{$J$} & \multicolumn{2}{c}{$g$}
    & \multicolumn{2}{c}{$q$ (CI)} & \multicolumn{2}{c}{$q$ (CI+$\Sigma$)} \\
      &   & \multicolumn{1}{c}{Expt.} & \multicolumn{1}{c}{Calc.}
    & \multicolumn{1}{c}{\VN} & \multicolumn{1}{c}{\VNm}
    & \multicolumn{1}{c}{\VN} & \multicolumn{1}{c}{\VNm} \\ 
\hline
$3d^4 4s\ ^6\!D$   & 5/2 & 1.669 & 1.657 & $-2483$ & $-2209$ & $-2430$ & $-2351$ \\
                   & 7/2 & 1.578 & 1.587 & $-2300$ & $-2034$ & $-2223$ & $-2145$ \\
$3d^4 4p\ ^6\!F^o$ & 1/2 &$-0.689$ &$-0.665$ & $-2052$ & $-1748$ & $-1979$ & $-1896$ \\
                   & 3/2 & 1.124 & 1.067 & $-1959$ & $-1661$ & $-1875$ & $-1792$ \\
                   & 5/2 & 1.314 & 1.314 & $-1807$ & $-1518$ & $-1705$ & $-1624$ \\
                   & 7/2 & 1.378 & 1.397 & $-1597$ & $-1321$ & $-1473$ & $-1395$ \\
                   & 9/2 & 1.416 & 1.434 & $-1333$ & $-1073$ & $-1180$ & $-1106$ \\
                   &11/2 &       & 1.454 & $-1016$ & $-776$ & $-830$ & $-758$ \\
$3d^4 4p\ ^6\!P^o$ & 3/2 & 2.382 & 2.385 & $-1607$ & $-1325$ & $-1489$ & $-1421$ \\
                   & 5/2 & 1.875 & 1.880 & $-1479$ & $-1209$ & $-1340$ & $-1280$ \\
                   & 7/2 & 1.710 & 1.714 & $-1281$ & $-1024$ & $-1117$ & $-1061$ \\
$3d^4 4p\ ^4\!P^o$ & 1/2 & 2.844 & 2.811 & $-2146$ & $-1782$ & $-2122$ & $-2003$ \\
                   & 3/2 & 1.802 & 1.786 & $-1913$ & $-1517$ & $-1847$ & $-1704$ \\
                   & 5/2 & 1.624 & 1.626 & $-1089$ & $-1036$ & $-916$ & $-946$ \\
$3d^4 4p\ ^6\!D^o$ & 1/2 & 3.155 & 3.186 & $-1512$ & $-1316$ & $-1390$ & $-1357$ \\
                   & 3/2 & 1.824 & 1.827 & $-1373$ & $-1227$ & $-1232$ & $-1232$ \\
                   & 5/2 & 1.628 & 1.634 & $-1651$ & $-1204$ & $-1547$ & $-1396$ \\
                   & 7/2 & 1.577 & 1.585 & $-1417$ & $-1171$ & $-1281$ & $-1220$ \\
                   & 9/2 & 1.570 & 1.552 & $-1200$ & $-977$ & $-1036$ & $-990$ \\
\end{tabular}
\end{ruledtabular}
\end{table}

Despite the consistency with respect to starting approximation, our transition energies still differ from experiment. The most likely explanation is that we haven't taken full account of the relaxation of the core $3p^6$ electrons. These have a strong effect on the $3d$ electrons via the exchange potential, yet relaxation of these orbitals is only taken into account to second-order using perturbation theory. Ideally one would include them as valence electrons in the CI so that their relaxation could be treated non-perturbatively, however this is not possible because the CI Hamiltonian size grows too large. In the \VNm\ approximation the $3p^6$ core is more tightly bound, lessening the magnitude of relaxation terms. This likely explains the improved agreement with experiment.

The $3p$ electrons also pose a potential problem for the isotope shift calculation. The scaled SMS operator that appears in \eref{eq:tilde_Q} manifests itself in the dipole part ($k=1$) of the multipole expansion of $\tilde Q$~\cite{berengut03pra}. Therefore it may be particularly affected if the \mbox{$3p$ -- $3d$} exchange terms are not adequately described by the method. It is for this reason that we conservatively use the difference between the CI and the CI+$\Sigma$ calculation as an estimate of the uncertainty in \ksms, rather than the smaller difference between the \VN\ and \VNm\ calculations.

We present our final values of relativistic shifts, $q$, for the $^6\!P^o$ transitions of astronomical interest (that is, those seen in quasar absorption spectra) in \Tref{tab:final_q}. As explained in \Sec{sec:results} we prefer our \VNm\ approximation which gives better agreement with experimental transition energies, however the difference between the \VN\ and \VNm\ calculations is similar to the errors quoted. The uncertainty is estimated as the difference between our pure CI calculation and the full CI+MBPT calculation including all $\Sigma$ diagrams. Our calculated $q$-values are seen to be in good agreement with the CI calculations of~\Cite{dzuba02praA}. In~\Tref{tab:qvalue} we see that the experimental $g$-factors for these transitions are well reproduced by our calculation. There are no close levels in this case, so the methods of matching $g$-factors~\cite{dzuba02praA} are not required. The $^4\!P^o_{5/2}$ and $^6\!D^o_{5/2}$ transitions are strongly mixed (as can be seen from the $g$-factors, which are well reproduced by our calculation), so if these transitions are ever seen in quasar absorption systems then a more careful analysis may be required.

\begin{table}[tb]
\caption{\label{tab:final_q} Relativistic shifts, $q$, for transitions to ground state \GS.}
\begin{ruledtabular}
\begin{tabular}{lcccbb}
Level  & $J$ & $\omega$ & $\lambda$ & \multicolumn{2}{c}{$q\ (\cm)$} \\
  && (\cm) & (\AA) & \multicolumn{1}{c}{this work} & \multicolumn{1}{c}{\Cite{dzuba02praA}} \\
\hline
$3d^4 4p\ ^6\!P^o$ & $3/2$ & 48399 & 2066 & -1421 (70) & -1360 (150)\\
                   & $5/2$ & 48491 & 2062 & -1280 (70) & -1280 (150)\\
                   & $7/2$ & 48632 & 2056 & -1061 (70) & -1110 (150)\\%
\end{tabular}
\end{ruledtabular}
\end{table}

In a previous comparison between theory and experiment in \SrII\ it was found that the SMS was underestimated by theory at the $\sim\!30\%$ level~\cite{lybarger11pra}. This is a single-valence-electron ion and so there is no CI; rather, we added \SigOne\ directly to our Dirac-Fock calculation. In the case of \SrII\ the addition of \SigOne\ did not adequately account for the effect of core-relaxation on the SMS. On the other hand, using the same method good agreement has been obtained between theory and experiment for \ZnII~\cite{berengut03pra}, which we also treated as a single-valence-electron ion. Moreover, because the majority of the isotope shift in \CrII\ comes from the valence-valence contributions which are treated to all orders using CI, we have good reason to believe that our mass-shifts have been calculated with reasonable accuracy.

We have also estimated the size of the field shift in these transitions using a small CI basis to estimate $F$ and experimental values of $\delta\!\left< r^2 \right>$ taken from~\cite{angeli04adndt}. The field shift is expected to be small for a light element like \CrII. Furthermore for the $d-p$ transitions of astrophysical interest the orbitals do not overlap the nucleus strongly and so there is additional suppression. We find the field shift is of order $\sim 0.010~\textrm{km}\,\textrm{s}^{-1}$ or smaller, which is much smaller than our uncertainty in \ksms. We neglect it entirely.

\begin{table}[tb]
\caption{\label{tab:final_IS} Calculated velocity structure in wavelength space of transitions to ground state \GS\ in \CrII.}
\begin{ruledtabular}
\begin{tabular}{lccbbb}
Upper level  & $J$ & $\lambda$ (\AA) & \multicolumn{3}{c}{$(\delta\lambda^{A,52}/\lambda) \cdot c$\ (km\,s$^{-1}$)} \\
 & & & \multicolumn{1}{c}{$A=50$} & \multicolumn{1}{c}{$A=53$} & \multicolumn{1}{c}{$A=54$} \\
\hline
$3d^4 4p\ ^6\!P^o$ & $3/2$ & 2066 & -0.535(96) & 0.252(45) & 0.495(89) \\
                   & $5/2$ & 2062 & -0.522(84) & 0.246(39) & 0.484(78) \\
                   & $7/2$ & 2056 & -0.535(96) & 0.252(45) & 0.495(89) \\%
\end{tabular}
\end{ruledtabular}
\end{table}

In \Tref{tab:final_IS} we present our isotope shift shift calculations for astronomically relevant transitions of \CrII\ using the \VNm\ results of \Tref{tab:sms} in~\eref{eq:IS}. Again the uncertainty quoted is the difference between the pure CI and CI+MBPT calculations, i.e. the entire effect of $\Sigma$. This is very much larger than the difference between our \VN\ and \VNm\ calculations. We quote the velocity structure in wavelength space relative to the leading isotope $^{52}$Cr. This is the preferred form for use in astronomy: the velocity shift is $\delta v = \lambda^{A,52}/\lambda \cdot c$ where $\lambda^{A,52}=\lambda^A - \lambda^{52}$ and $c$ is the speed of light in km\,s$^{-1}$.

\section{Conclusion}

We have shown that the CI+MBPT method can give transition energies in good agreement with experiment for low-lying transitions in \CrII. Although the subtraction diagrams are very large when the orbitals are calculated in the \VNm\ or \VN\ approximations, when all second-order MBPT diagrams are taken into account consistently the calculated energies are found to improve. This may help to direct future efforts using the CI+MBPT method in many-valence-electron ions such as \FeII, which is of importance to studies of $\alpha$-variation in quasar absorption systems.

The SMS in \CrII\ is found to dominate the isotope shift. For the $\GS \rightarrow 3d^4 4p\ ^6\!P^o$ transitions seen in quasar spectra they are five times the magnitude of the normal mass-shift and of opposite sign. The total mass-shift constant for these transitions, $\knms+\ksms = 3365\,(112)$~\GHzamu\ (taking the $J=3/2$ upper level), is consistent with an earlier CI estimate of $1900\,(1200)$~\GHzamu~\cite{kozlov04pra}, although clearly at the limit of the uncertainty. The \CrII\ isotope shift is also quite large in comparison to many of the other isotope shifts used in the quasar analysis (although many are unknown). The velocity shift between the even isotopes ($\sim 500\,\ms$) is comparable to the isotope shifts of the $\lambda\lambda 2803$ and 2796 lines of \MgII\ ($\sim 850\,\ms$). Fortunately in the case of \CrII\ there are stable isotopes on either side of the leading isotope, so one may hope that the total systematic shift due to variation of isotope abundances is small.

\acknowledgments

I thank Victor Flambaum and Michael Murphy for useful discussions. This work was supported by an award under the Merit Allocation Scheme on the NCI National Facility at the ANU and by the Australian Research Council (DP110100866).

\bibliography{references}

\end{document}

%% file: phys_jdf.tex
\def\apj{Astrophys. J.}

\def\adndt{At. Data Nucl. Data Tables}

\def\hypint{Hyp. Int.}

\def\jpb{J. Phys. B}

\def\jpcs{J. Phys.: Conf. Ser.}

\def\lnp{Lect. Notes Phys.}

\def\mnras{Mon. Not. R. Astron. Soc.}

\def\pra{Phys. Rev. A}

\def\prd{Phys. Rev. D}

\def\prl{Phys. Rev. Lett.}